\documentclass[11pt, oneside]{article}   	
\usepackage{geometry}                		
\usepackage{graphicx}				
\usepackage{amssymb}
\usepackage{authblk}
\usepackage{natbib}
\usepackage{url}

\title{An agent-based model of citation behavior}

\author[1]{George Chacko \footnote{chackoge@illinois.edu}}
\author[2]{Minhyuk Park\footnote{Contributed equally.}}
\author[3]{Vikram Ramavarapu\footnote{Contributed equally.}}
\author[4]{Ananth Grama}
\author[5]{Pablo Robles Granda}
\author[6]{Tandy Warnow}
\affil[1,2,3,5,6,]{Siebel School of Computing and Data Science, Grainger College of Engineering, University of Illinois Urbana-Champaign, IL, USA}
\affil[4]{Department of Computer Science, Purdue University}

\begin{document}
\maketitle 



\begin{abstract}
Whether citations can be objectively and reliably used to measure productivity and scientific quality of articles and researchers can, and should, be vigorously questioned. However, citations are widely used to estimate the productivity of researchers and institutions, effectively creating a `grubby' motivation to be well-cited. We model citation growth, and this grubby interest using an agent-based model (ABM) of network growth. In this model, each new node (article) in a citation network is an autonomous agent that cites other nodes based on a `citation personality' consisting of a composite bias for locality, preferential attachment, recency, and fitness.  We ask whether strategic citation behavior (reference selection) by the author of a scientific article can boost subsequent citations to it.  Our study suggests that fitness and, to a lesser extent, out\_degree and locality effects are influential in capturing citations, which raises questions about similar effects in the real world. \\

\emph{keywords}: agents; mid-level models; citation networks

\end{abstract}




\newpage

\section{Introduction} 

Interest in a theory of citation has generated models, discussion, disagreement, proposed refinements and re-interpretations \citep{Price1965,Chubin1975,Moravcsik1975,Price1976,Gilbert1977,Small1978,Cozzens1981,Zuckerman1987,Martin1983,MacRoberts1996,Stigler8bba9ab5c485,Leydesdorff1998,Wouters1999TheCC,Aksnes2019,Olszewski2020}. A unifying theory is not evident and uncertainty remains around the basis for citation behavior, although information retrieval \citep{Cronin1981TheNF}, rhetorical \citep{Gilbert1977,Cozzens1989}, and reward-based motivations \citep{Merton1968,Latour1986-sy} likely act in concert with wide variation across individuals, communities, and disciplines.

 Theoretical uncertainty notwithstanding, citation analysis is heavily used to evaluate the productivity and performance of researchers. An effect of such evaluations is to trigger an interest in being well-cited, noted years ago in `The Citation-Index Game' \citep{Leopold1973}. This interest may reasonably be grouped with other "grubby'' motivations of scientists \citep{Hacking1994,Kitcher1993-KITTAO-2}, yet  is not inconsistent with extant theories of citation. For example, wanting to be well-cited may, for example, also be implicit in an author's expectation that an article is (i) rewarded for its scholarly contribution, (ii) cited on account of being persuasive.

Quantitatively analyzing observed behavior and reconciling it with motivations that are qualitatively described is challenging. However, it is possible to examine citation patterns and use them to explain resultant citation landscapes. In this respect, empirical perspectives  are useful and agent-based modeling offers value through stochastic simulations under idealized models. We agree with  \cite{Watts2011}, who argue that simulations occupy an important space between a replication of the process being studied and an abstract mathematical model. We have designed and implemented an agent-based model (ABM) that models the growth of citation networks through a combination of generative processes and allows us to simulate the citation process as an article is published.

Our model enables simulations of the growth of citation networks by adding new nodes to an existing citation graph consisting of articles (nodes) linked by citations. The use of an idealized model also reduces concerns relating to data quality and sampling error \citep{MacRoberts1996}.  We approach citations from the perspective of strategy and ask the narrow question of whether the choice of references made by the author of an article can influence the number of citations it receives, and more generally examine how citation behavior affects the growth of citation networks. Within the framework of this question target node properties, individual agent behavior, and the network environment contribute to the simulation process. 

The model is relatively simple. In a vertex-copying approach \citep{Kleinberg1999,Krapivsky2005}, an agent is generated by cloning an existing node (generator).  A newly created agent always cites its generator, and  a parameter, $\alpha$, determines the fraction of citations that are made  within  the neighborhood of the generator node. Each year a number of new nodes (agents) are created. When an agent is generated, its initial citation count is set to zero and it is assigned a fitness value  \citep{Touwen2024,Zhou2023} reflective of quality. In our model, fitness is presently drawn from a power-law distribution and does not change over time (Materials and Methods).
 The agent is also assigned a quota of references it can make by random selection from a distribution derived from real-world citation data.

Each agent has a phenotype, which expresses user-defined bias towards fitness, preferential attachment \citep{simon1955class,Price1976,barabasi1999emergence}, immediacy or recency \citep{Price1965,Silva2020}, and community context \citep{Gilbert1997,Wedell2022} when choosing which nodes to cite.  Each agent  then picks other nodes to cite, using a weighted sampling function based on the agent's phenotype. Simulations are conducted in time-steps of one year.  
Note that the agent's phenotype determines the relative importance of fitness, in-degree (which determines the preferential attachment score), year of publication, and whether a node is in its community.  
For example, when fitness and preferential attachment are equally weighted, then the assigned fitness of a node drives the early period of citation accumulation, and gradually decreases as the node accumulates in-degree.  

A large body of interdisciplinary research precedes this article that, if discussed, would render this Introduction tediously long. Therefore, we restrict ourselves to mentioning three more relevant efforts in addition to those already cited above. \cite{Borner2004} developed a citation model that included an author variable that was validated against observed against data from a single journal,  in which they discussed recency and topics as influences on preferential attachment mechanisms. \cite{,Watts2011} performed an agent-based simulation that considered a fitness landscape when modeling the generation of articles also validated against data from a single journal. \cite{Shah2019} described a random walk method for network generation that accounted for bounded rationality, homophily, triadic closure, locality, and preferential attachment.

Our technical approach combines ideas from the work of \cite{simon1955class}, \cite{Touwen2024},  \cite{Gilbert1997}, and \cite{Weisberg2009}. The latter two used a planar epistemic grid as a starting point for simulations of the structure of academic science and the division of cognitive labor respectively. Presuming richer epistemic connections, we opt to use a citation network as a starting point.  We restrict the realm of possibility by adopting a mid-level modeling approach \citep{Harnagel2019,Alexandrova2017} that uses parameter settings derived from real-world data in order to focus on likely outcomes rather than all possible outcomes.  In this article, we use citation data concerning the emergence of the field of exosome and extracellular vesicle biology \citep{Raposo2023}, although the implementation of our model makes it possible to use parameters drawn from other citation graphs.

Under simulation conditions where the agent environment is either homogeneous or heterogeneous with respect to agent phenotypes, we ask whether some parameter settings offer an advantage in accumulating citations. In the following sections, we describe our model in greater detail, show results from simulations and discuss our findings. We acknowledge the limitations of idealized models, and reiterate that any conclusions drawn are relevant to the artificial world of an ABM, although the results pose questions that can be asked of the real world. Open source code that we developed for this purpose is freely available and is actively being improved for extensibility and scalability.

\section{Results}

In this section, we present results from baseline observations and the effects of introducing agents with high fitness (superstar nodes), varying out\_degree, and locality.

 \subsection{Baseline Observations} 

 In an initial assessment of our model, we compared citation counts per node generated in different environments. Simulations were conducted under conditions where the agent environment was either heterogeneous (randomized or \emph{ra}) or homogeneous (static or \emph{sa}) with respect to agent phenotype. For \emph{ra}, every agent was assigned values of $\alpha$ between 0 and 1, and values for preferential attachment, recency, and 
 fitness that summed to 1. For \emph{sa}, every agent was instantiated with $\alpha$=0.5 and equal bias towards preferential attachment (1/3),  recency (1/3), and fitness (1/3).  Every agent was assigned a phenotype, i.e., a linear combination of weights for preferential attachment, recency, and fitness (Materials and Methods). 
 
To seed each simulation, we used two datasets: either the real-world derived Stahl-Johnstone citation graph (\emph{sj}) or an  Erd\H{o}s-R\'{e}nyi random graph (\emph{er}) with the same density, described in Materials and Methods.  All simulations consisted of 30 cycles at a 3\% annual rate of growth, referred to as `standard' conditions in the rest of this manuscript. Starting from a seed set of $\sim$ 492,000 nodes, these conditions resulted in a network  of  approximately 1,193, 000 nodes, due to the generation of about 700,000 agents. 
 
 Three replicates were conducted for each of the four conditions (\emph{ra}, \emph{sa}, \emph{sj}, \emph{er}).  Median global and average local clustering coefficients for the three replicates are shown in Table \ref{tab:gcc_lcc} and indicate a slightly higher global clustering coefficient for networks initiated with an \emph{er} graph and slightly higher mean local clustering coefficient for simulations run in a static agent background, relative to those involving agents with randomly assigned phenotypes.

Six simulations were performed on each dataset (Table~\ref{tab:table1-posit-meas}). For \emph{sj} data, the initial graph of 491,532 nodes and 899,050 edges was transformed into a graph of 1,193,102 nodes and between 14,581,243 and 14,647,070 edges, with a median value of 14,634,098 edges. 

The  \emph{er} graphs have isolated nodes (an expected outcome given their density), which makes them different from the \emph{sj} graphs, which have no isolated nodes.  
Therefore, we performed simulations on three different \emph{er} graphs. The output from simulations on \emph{er} graphs consisted of between 1,162,054 and 1,162,746 nodes and edge count ranging between 14,240,417 and 14,297,067. The differences in median values for nodes and edges across the two groups (\emph{sj} and  \emph{er}) were in the order of 3\% and .03\% respectively and were considered unlikely to confound comparisons. 

\begin{figure}[!t]
\centering
\includegraphics[width=0.8\textwidth]{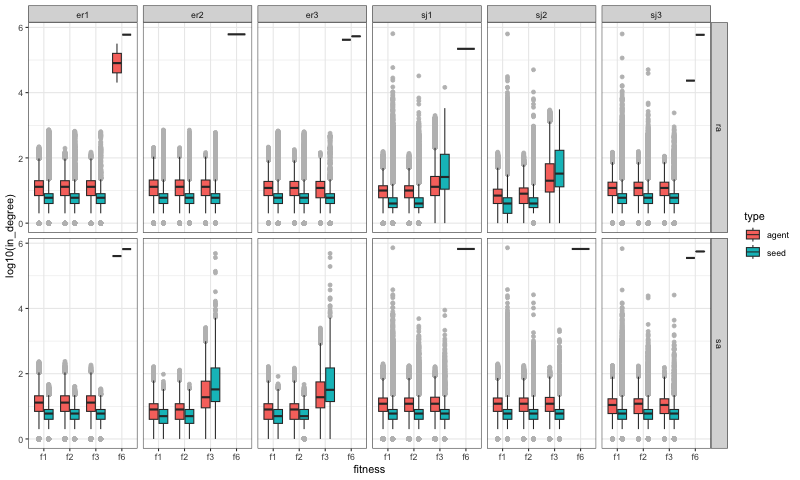}
\caption{\emph{Citation counts of nodes (in\_degree) under default simulation conditions} Using either the Stahl-Johnstone real-world \emph{sj} dataset or Erd\H{o}s-R\'{e}nyi (\emph{er}) graphs as input, simulations were performed in either random (\emph{ra}) or homogenous (\emph{sa}) agent backgrounds for 30 cycles of 3\% annual growth (standard), resulting in an increase from roughly half a million nodes to 1.2 million nodes. In\_degree of nodes is plotted against fitness groups using a log10 scale on the y-axis for in\_degree after excluding agents with zero in\_degree (roughly 4.6\% for \emph{sj} and 4.2\% for \emph{er}). Nodes are grouped by fitness values: f1(1:10), f2(10:100), f3(100:1,000), f4(1,000:10,000), f5(10,000:100,000), f6(100,000: 1,000,000). Due to the power law distribution of fitness, roughly $85$\% of the observations in each simulation are found in group f1, $12$\% in f2, $3$\% in f3 and $\sim$~$10^{-5}$\% in f6.} 
\label{fig:er_sj_comp}
\end{figure}

Variation was seen across \emph{ra} and \emph{sa} agent backgrounds, across starting point data (\emph{sj} or \emph{er} datasets) and within replicates (Figure  \ref{fig:er_sj_comp}).  For ease of viewing, agents were grouped into six fitness groups of progressively increasing fitness (Figure \ref{fig:er_sj_comp}). Within each group, the median citation count of agents was greater than the median citation count of corresponding seed nodes in 8 of 12 cases. Exceptions seen in Figure \ref{fig:er_sj_comp} concern fitness group 3 in 4 of 12 simulations. We speculate that these are cases where higher fitness values override  recency effects, since agents are time-stamped with later years than seed nodes. 


\begin{table}[ht]
\caption{\emph{Clustering Coefficients.} Global (gcc) and average local clustering coefficients (alcc) were computed for median values of three replicates each of 30 year and 3\% per year simulations, initiated with either the \emph{sj} or \emph{er} in an agent background of either random (\emph{ra}) or static agent environments (\emph{sa}).}
\centering
\vspace{2 mm}
\begin{tabular}{rllrr}
  \hline
 & dataset & agent\_bg  & gcc & alcc \\ 
  \hline
1 & er & ra & 0.0001874 & 0.1664000 \\ 
  2 & er & sa & 0.0001837 & 0.1988000 \\ 
  3 & sj & ra & 0.0001398 & 0.1741000 \\ 
  4 & sj & sa & 0.0001306 & 0.1981000 \\ 
   \hline
\end{tabular}
\label{tab:gcc_lcc}
\end{table}

However,  seed nodes exhibited a greater number of high in\_degree outliers  (Figure  \ref{fig:er_sj_comp}), which we attribute to the number of nodes in the \emph{sj} dataset with already high in\_degree that become a target for agents biased towards preferential attachment. For example, the {sj} dataset had 6,263 nodes with in\_degree $> 10$ and 60 nodes of in\_degree $> 1000$. In comparison, the Erd\H{o}s-R\'{e}nyi graph (\emph{er1}) has 102 nodes with in\_degree $> 10$ and none of  in\_degree $> 1000$.

In all cases, where agents had  high in\_degree of around 500,000 (Table \ref{tab:table1-posit-meas}), they were of correspondingly high fitness. For the \emph{er} replicates, a heterogenous (\emph{ra})  agent background increased the number of relatively high in\_degree outliers compared to the static background (\emph{sa}). We discuss the relationship between fitness, in\_degree, and age of the agent further below.

These data suggest that fitness is the major driver of citation, that high in\_degree at the beginning of a simulation confers an advantage via preferential attachment, and that recency is influential  for nodes initialized with low in\_degree. The read-out of in\_degree does not suggest any advantage to using a citation graph over a random graph as a starting point. If epistemic richness, which we presume is encoded in the \emph{sj} dataset, exerts influences on the outcome of the simulation, such influence is not detected by a simple measurement of accumulated in\_degree.

\begin{table}[ht]
\caption{Positional statistics of in\_degree from simulations shown in Figure \ref{fig:er_sj_comp}. Data represent both seed and agent nodes. q0.9 and q0.99 denote the 90th and 99th percentile of in\_degree respectively. sj\_rep0, sj\_rep0, and sj\_rep2 are three replicates of simulations on the \emph{sj} dataset and er1, er2, and er3 represent a single simulation each of three Erd\H{o}s-R\'{e}nyi graphs. \emph{ra} and  \emph{sa} refer to randomized and homogeneous agent environments respectively}
\centering
\begin{tabular}{rlrrrrr}
  \hline
 & tag & min & med & q0.9 & q0.99 & max \\ 
  \hline
1 & er1\_ra & 1 & 8 & 24 & 52 & 598,318 \\ 
  2 & er1\_sa & 1 & 8 & 25 & 52 & 658,608 \\ 
  3 & er2\_ra & 1 & 8 & 25 & 53 & 612,102 \\ 
  4 & er2\_sa & 1 & 6 & 16 & 58 & 482,948 \\ 
  5 & er3\_ra & 1 & 8 & 23 & 49 & 533,129 \\ 
  6 & er3\_sa & 1 & 6 & 16 & 58 & 485,801 \\ 
  7 & sj\_rep0\_ra & 1 & 7 & 16 & 75 & 640,947 \\ 
  8 & sj\_rep0\_sa & 1 & 8 & 21 & 49 & 724,489 \\ 
  9 & sj\_rep1\_ra & 1 & 5 & 14 & 93 & 630,363 \\ 
  10 & sj\_rep1\_sa & 1 & 8 & 21 & 49 & 723,807 \\ 
  11 & sj\_rep2\_ra & 1 & 8 & 21 & 50 & 628,743 \\ 
  12 & sj\_rep2\_sa & 1 & 8 & 20 & 47 & 679,832 \\ 
   \hline
\end{tabular}

\label{tab:table1-posit-meas}
\end{table}

\newpage

\subsection{``Superstar'' Effects}  

A few articles in the literature, often describing methods,  have been heavily cited for extended periods of time, with a notable example being \cite{Lowry1951}, which has over 300,000 citations today  \citep{VanNoorden2014}. \cite{Lowry1951} is one of the nodes in the \emph{sj} dataset where it has 88,435 citations in-graph. We added the ability to introduce a small number of such nodes into our simulation by assigning high fitness values to them. Specifically, we planted one node each with fitness levels of 10,000, 100,000, and 1,000,000 respectively at the beginning of each simulation. These fitness levels correspond to 10, 100, and 1,000 times greater than the maximum fitness levels assigned by our power-law method.

\begin{figure}[!t]
\centering
\includegraphics[width=0.7\textwidth]{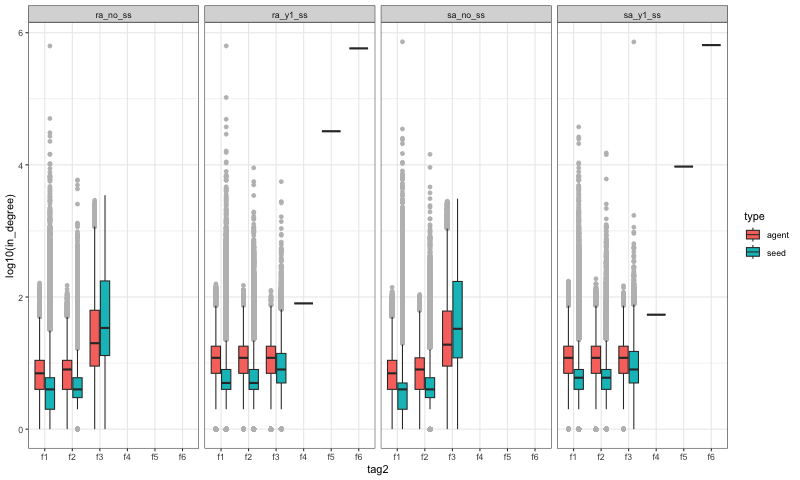}
\caption{\emph{Effect of Superstar Nodes} Using real-world \emph{sj} data as input, standard simulations were performed in either random (\emph{ra}) or homogenous (\emph{sa}) agent backgrounds resulting in an increase from roughly half a million nodes to 1.2 million nodes. To examine the effects of high fitness nodes, three `superstar' agents with fitness 10,000, 100,000, and 1,000,000 were generated in year 1 of the simulation, allowing 29 years in which to accumulate citations (fitness groups 4-6, panels 2 and 4 from left to right). A reduction of in\_degree is seen in f3 when superstars are present irrespective of whether the background is random or static. Nodes are grouped by fitness values f1(1:10), f2(10:100), f3(100:1,000), f4(1,000:10,000), f5(10,000:100,000), f6(100,000: 1,000,000). A log10 scale is used on the y-axis for in\_degree after excluding agents without any in\_degree, (roughly 4.6\% for \emph{sj} and 4.2\% for \emph{er}). The control group (no\_ss) does not have any agents with fitness $> 1,000$.} 
\label{fig:pl_ss}
\end{figure}

Results in Figure \ref{fig:pl_ss}  show that a small number of planted nodes of high fitness (0.00043\% of agents and 0.00025\% of total nodes) accumulate citations proportional to their fitness values. Interestingly, a `quenching effect' is seen on other nodes, especially evident in fitness group 3, which exhibits a noticeable reduction in the median and quartile values of in\_degree for agent nodes compared to the control group without planted high-fitness nodes. Also evident is the effect on seed nodes, which we tentatively interpret as fitness overriding in\_degree. We note, however, that our simulation approach effectively imposed `zero sum game' conditions, which must be considered even while raising the question of whether competition for citations also occurs in real-world situations. Exploring such counterfactual scenarios may not be trivial.

\newpage

\subsection{Effect of out\_degree} 

\begin{figure}[!t]
\centering
\includegraphics[width=0.6\textwidth]{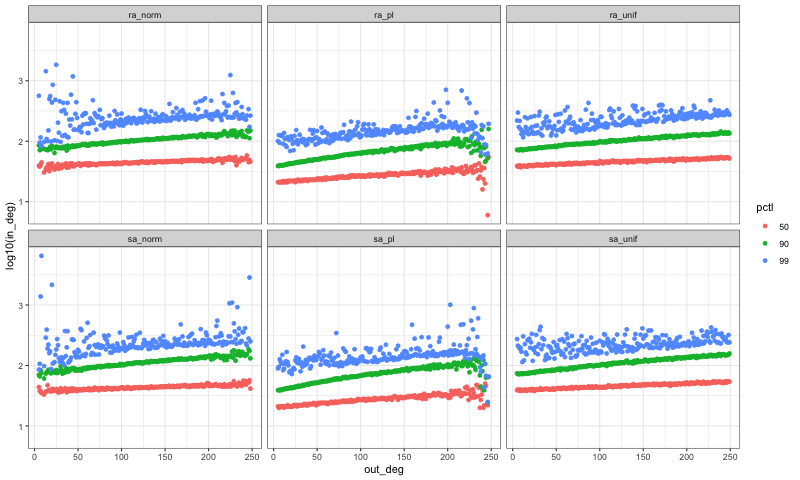}
\caption{\emph{Reference Count Effects} Higher out\_degree is associated with higher in\_degree for power-law, normal, and uniform distributions of out\_degree. Standard simulations were conducted on the \emph{sj} dataset in either \emph{ra} or  \emph{sa} agent backgrounds. For the \emph{ra} background, $\alpha$ was randomized from 0 to1 and weights for preferential attachment, recency and fitness were randomly assigned that summed to 1. For \emph{sa},  $\alpha$ was set to 0.5 and weights for preferential attachment, recency and fitness were set to $0.33\bar{3}$,  summing to 1. Out\_degree was assigned to agents from real-world values ranging from 5-249 that were fit to  power-law, normal, and uniform distributions (Materials and Methods). Median, 90th, and 99th percentile values of in\_degree (pctl) are plotted against out\_degree.}
\label{fig:out_deg_effect}
\end{figure}

Previous observations on fractional counting of articles and citations \citep{DeSollaPrice1966,Small1985sw,Leydesdorff2010} stimulated us to ask the following question in the context of our simulations: whether the number of references in a node (out\_degree) can influence the number of citations (in\_degree) it receives (Figure \ref{fig:out_deg_effect}). 

\cite{Castillo-Castillo2025-ov} argue for out\_degree being a power-law distribution. However, this was not especially obvious in our samples of real-world networks. Therefore, we generated out\_degree distributions limited to the range of values seen in our real-world data according to normal, power-law, and uniform distributions.  

The results show that agents with higher reference counts (out\_degree) tend to be associated with higher citation counts (in\_degree), whether the agent background is either randomized or static. Within the confines of a simulation, this experiment result suggests a cynical strategy towards being well-cited by increasing the number of references in a manuscript. Whether the strategy would work in the real world is much challenging to test but the result raises the question of whether well-cited articles are being acknowledged not only because of their content but also because of the number of references cited.

\newpage
\subsection{Alpha Effects}

The $\alpha$ parameter controls locality effects. When $\alpha$  is set to zero, all citations made by an agent are made to nodes outside the neighborhood of its generator (Introduction). When the $\alpha$  parameter is set to 1, all citations made by an agent are made to nodes within its inherited neighborhood and only unused citations are carried over to nodes outside the neighborhood of its generator. An $\alpha$ setting of 0.5, splits citations equally between within and outside the neighborhood of its generator: the neighborhood it `inherits' from its generator.

\begin{table}[ht]
\caption{Global clustering (gcc) and average local clustering (lcc) coefficients as $\alpha$ is varied}
\centering
\begin{tabular}{rlllrr}
  \hline
 & alpha & bg & ss & gcc & lcc \\ 
  \hline
  1 & alpha\_0.0 & ra & no\_ss & 0.000093 & 0.033619 \\ 
  2 & alpha\_0.0 & ra & ss & 0.000107 & 0.049218 \\ 
  3 & alpha\_0.0 & sa & no\_ss & 0.000088 & 0.038677 \\ 
  4 & alpha\_0.0 & sa & ss & 0.000110 & 0.051326 \\ 
  \hline
  5 & alpha\_0.5 & ra & no\_ss & 0.000138 & 0.201420 \\ 
  6 & alpha\_0.5 & ra & ss & 0.000132 & 0.201735 \\ 
  7 & alpha\_0.5 & sa & no\_ss & 0.000128 & 0.209489 \\ 
  8 & alpha\_0.5 & sa & ss & 0.000130 & 0.199147 \\ 
  \hline
  9 & alpha\_1.0 & ra & no\_ss & 0.000208 & 0.251987 \\ 
  10 & alpha\_1.0 & ra & ss & 0.000198 & 0.231878 \\ 
  11 & alpha\_1.0 & ss & no\_ss & 0.000187 & 0.258471 \\ 
  12 & alpha\_1.0 & ss & ss & 0.000200 & 0.228021 \\ 
   \hline
\end{tabular}

\label{tab:alpha_gcc_lcc}
\end{table}

\begin{table}[ht]
\centering
\caption{Clustering. When $\alpha>0$, static agent backgrounds form large clusters. Networks were generated varying $\alpha$, agent background (\emph{bg}), and +/- planted superstar agents of as below to form a set of 12 networks. Each network was clustered using IKC (a recursive k-core clustering technique)   \emph{nc}: node coverage--the percentage of nodes in clusters of at least size two. \emph{cc}: count of clusters generated. \emph{cs}: cluster size, \emph{k}: value of k. Singleton clusters were not considered in the descriptive statistics below. The experiment was repeated twice more to generate 36 networks in total, and median values across replicates are shown.}
\vspace{1mm}
\begin{tabular}{rlrrrrrrr}
  \hline
  & $\alpha$ & bg & ss & median(nc) & median(cc) & median(cs) & median(k)  \\ 
  \hline
1 & 0.0 & ra & no\_ss & 27.60 & 27.00 & 79.50 & 24.00 \\ 
  2 & 0.0 & ra & ss & 33.90 & 28.00 & 163.00 & 61.50 \\ 
  3 & 0.0 & sa  & no\_ss & 2.73 & 6.00 & 131.00 & 10.50 \\ 
  4 & 0.0 & sa & ss & 31.47 & 5.00 & 90.00 & 10.00 \\ 
  \hline
  5 & 0.5 & ra & no\_ss & 26.01 & 25.00 & 65.50 & 21.00 \\ 
  6 & 0.5 & ra & ss & 20.91 & 18.00 & 251.50 & 56.00 \\ 
  7 & 0.5 & sa & no\_ss & 9.12 & 2.00 & 54,319.00 & 33.00 \\ 
  8 & 0.5 & sa & ss & 22.62 & 1.00 & 269,473.00 & 20.00 \\ 
  \hline
  9 & 1.0 & ra & no\_ss & 20.92 & 24.00 & 71.00 & 22.00 \\ 
  10 & 1.0 &  ra & ss & 22.58 & 22.00 & 213.00 & 58.50 \\ 
  11 & 1.0 &  sa & no\_ss & 11.25 & 3.00 & 12,103.00 & 14.00 \\ 
  12 & 1.0 & sa & ss & 16.04 & 2.00 & 107,095.00 & 21.00 \\ 
   \hline
\end{tabular}
\label{tab:ikc_data}
\end{table}

Increasing $\alpha$ in a simulation results in higher global and average local clustering coefficients (Table \ref{tab:alpha_gcc_lcc}), which are measures that reflect, to some extent, how nodes in a network tend to cluster. These trends suggest that $\alpha$ can be used as a parameter to influence community structure. We also evaluated community structure by clustering the networks using the IKC algorithm \citep{Wedell2022}, which recursively extracts k-cores from a network beginning with the densest (Table \ref{tab:ikc_data}). For \emph{ra} networks, node coverage (the fraction of nodes in clusters of at least size two) did not change significantly as $\alpha$ varied between 0, 0.5 and 1, whether superstar nodes were present or absent. For the static agent background (\emph{sa}), however, node coverage increased roughly tenfold as superstar nodes were introduced. 
The number of clusters for \emph{sa} was low relative to \emph{ra},  while median cluster size was large relative to the \emph{ra} background. The presence of superstar nodes for \emph{sa} background resulted in a roughly five-fold and ten-fold increase respectively in median cluster sizes when $\alpha$ was 0.5 and 1.0. For the \emph{ra} background a five-fold increase in median cluster size is also seen but median cluster sizes are much smaller and the number of clusters is greater. Collectively, the data suggest that \emph{ra} and \emph{sa} agent background exhibit differences in community structure with the former tending to form smaller clusters while both backgrounds are affected by the presence of superstar nodes. Of the two, \emph{ra} appears more realistic  given that one would expect a variety of behaviors in the scientific world.

\begin{figure}[!t]
\centering
\includegraphics[width=0.8\textwidth]{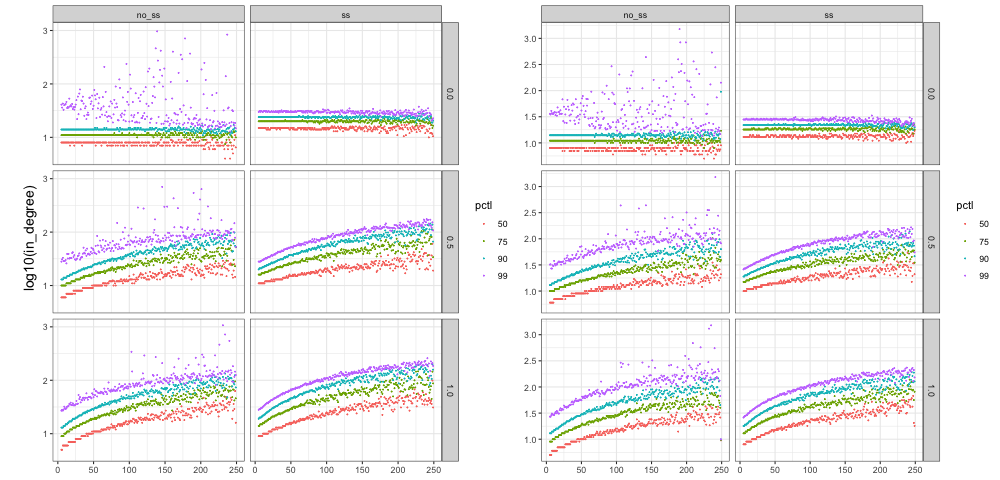}
\caption{\emph{Reference Count Effects as $\alpha$ is Varied}  The effect of out\_degree on in\_degree is modulated by  $\alpha$. For $\alpha >0$, higher out\_degree (x-axis) is associated with higher in\_degree (y-axis) for real-world derived distributions of out\_degree; in contrast, there is no impact when $\alpha=0$. Median, 75th, 90th, and 99th percentile values of agent in\_degree (pctl) are plotted against out\_degree. Standard simulations were conducted on the \emph{sj} dataset in either \emph{ra} or \emph{sa} agent backgrounds. Left panel: \emph{ra} Right panel: \emph{sa}}
\label{fig:simplified_extreme_alphas}
\end{figure}

 One prediction is that increasing $\alpha$ should increase the positive correlation between  out\_degree and indegree seen in Figure \ref{fig:out_deg_effect}. We varied $\alpha$ from 0 to 1,  in steps of 0.5, and observed an increase in the values of the 90th and 99th percentiles of in\_degree (Table \ref{tab:alpha_effects}). We also repeated the experiment reported in Figure \ref{fig:out_deg_effect},  while additionally varying $\alpha$ and using the observed frequencies of out\_degree in the \emph{sj} dataset.  We observed a dose-dependent effect (Figure \ref{fig:simplified_extreme_alphas}). These data suggest a mechanism by which higher out\_degrees have a local effect at higher values of $\alpha$ by restricting citations to a smaller number of nodes. However, maximum in\_degree was observed when $\alpha$ was set at 0.5 in the presence of superstar nodes (Table \ref{tab:alpha_effects}).

\begin{table}[ht]

\caption{Positional statistics of Agent in\_degree as $\alpha$ is varied}
\centering
\begin{tabular}{rlllrrrrrr}
  \hline
 & alpha & bg & ss & min & median & q0.75 & q0.90 & q0.99 & max \\ 
  \hline
1 & alpha\_0 & ra & no\_ss & 0.00 & 8.00 & 11.00 & 14.00 & 37.00 & 3,023 \\ 
  2 & alpha\_0 & ra & ss & 0.00 & 13.00 & 18.00 & 22.00 & 28.00 & 464,316 \\ 
  3 & alpha\_0 & sa & no\_ss & 0.00 & 8.00 & 11.00 & 14.00 & 40.00 & 3,177 \\ 
  4 & alpha\_0 & sa & ss & 0.00 & 15.00 & 20.00 & 24.00 & 30.00 & 487,226 \\ 
  \hline
  5 & alpha\_0.5 & ra & no\_ss & 0.00 & 7.00 & 11.00 & 16.00 & 53.00 & 2,676 \\ 
  6 & alpha\_0.5 & ra & ss & 0.00 & 11.00 & 17.00 & 23.00 & 44.00 & 615,937 \\ 
  7 & alpha\_0.5 & sa & no\_ss & 0.00 & 7.00 & 11.00 & 16.00 & 50.00 & 2,813 \\ 
  8 & alpha\_0.5 & sa & ss & 0.00 & 12.00 & 18.00 & 24.00 & 48.00 & 649,547 \\ 
  \hline
  9 & alpha\_1.0 & ra & no\_ss & 0.00 & 6.00 & 11.00 & 17.00 & 61.00 & 2,532 \\ 
  10 & alpha\_1.0 & ra & ss & 0.00 & 9.00 & 16.00 & 24.00 & 58.00 & 457,899 \\ 
  11 & alpha\_1.0 & sa & no\_ss & 0.00 & 6.00 & 11.00 & 17.00 & 60.00 & 2,663 \\ 
  12 & alpha\_1.0 & sa & ss & 0.00 & 10.00 & 17.00 & 26.00 & 63.00 & 480,383 \\ 
   \hline
\end{tabular}
\label{tab:alpha_effects}
\end{table}

\newpage
\section{Summary and Conclusions} 

We have developed software for agent-based simulations that permits simulation of referencing and citation behavior inherent in the scientific publication process. We report initial findings from simulations of a simple model that considers preferential attachment, recency, and fitness mechanisms. The study opens up a number of questions. 

Under our conditions of simulation, we find that fitness is a major driver of citations. This observation suggests that intrinsic quality may be a dominant influence, despite other motivations that have been propose such as rhetorical. However, extended simulation studies supported by corresponding findings in real-world data will be necessary to firm up this tentative conclusion. For example, the power-law distribution of fitness we use may need to be tuned or replaced by other distributions. Additionally, it is worth studying the case where fitness decays over time.

The mid-level model we used applies constraints derived from publications in cell biology. These constraints are largely derived from a single citation graph of around 14 million nodes  that was generated by a seed set expansion protocol. Whether the trends seen would change significantly with graphs derived from different disciplines and drawn from different time periods remains to be ascertained.

The superstar effects we observed resulted in a quenching effect on the in\_degree of some agents and seems to increase the citations of others with lower fitness. Whether this quenching phenomenon occurs in the real world, and under what conditions, remains to be ascertained. We have not discussed subtle trends in our observations since they require further examination but it is possible that they may be of interest to those who compete ardently for citations and find appeal even in small increases in citation counts.

Interestingly,  when $\alpha>0$, increasing the number of references in an article results in more citations. We attribute this to an indirect community structure effect that results in a greater probability of being cited, which is supported by results from varying the $\alpha$ parameter.  At this point in time, we do not have any conclusive evidence from seeding a random selection of phenotypes into a static background results in any particular phenotype being most highly cited. 
In contrast, when $\alpha=0$ we do not see these effects. The contrast between $\alpha=0$ (where there is no community allegiance) and $\alpha>0$ (where there is community allegiance) is noteworthy.  Given the expectation that researchers do identify with some community, the setting of $\alpha>0$ is more likely to be representative of real-world behavior.

In this initial model, we make the simplifying assumption that every article is written by a single author and no author has written more than one article. Extending the model to handle authorship in a more realistic manner is planned, which would also enable modeling self-citations and aspects of post-epistemic misconduct involving citations \citep{biagioli2020}. Another future direction is to model the evolution of research communities from a founder or small set of founder publications. 

Our software is being engineered for scalability and performance with the goal of being able to grow a network from a single node to around 150 million nodes with reasonable cost. In other words, to be able to simulate the evolution of the scientific literature from a single article to its present state. Scaling will involve further algorithmic development and specialized hardware configurations, since computing scores for all nodes in each cycle becomes expensive as the network grows. Last, we intend to use our software alone and in combination with other generators to construct synthetic networks.

\section{Materials And Methods}

\subsection {Overview} The simulation process consists of agents sequentially created in batches and allowed to make citations. The input to a simulation is a real-world or random graph and a set of parameters. The output is a generated synthetic graph. During a simulation, agents are created in batches corresponding to a growth rate for a year. An agent is a representation of an article written by an author. Each new (agent) is generated by randomly selecting any existing node, also termed a generator node, and cloning it as an agent. This agent cites its generator and also `inherits' the neighborhood of its generator. The agent is assigned a quota of references by random selection from a distribution derived from real-world data. The agent is assigned a user-specified phenotype: a bias for target selection based on in-degree, year of publication, fitness, and locality. The phenotype consists of three corresponding weights for preferential attachment, recency, and fitness respectively that drive target selection (below).

Locality is defined by the parameter ($\alpha$), which determines the fraction of an agent's references that cite nodes in the neighborhood of its generator node. Thus, if an agent is assigned 30 references and an $\alpha$ value of 0.5, it will make 1 reference to its generator and 14 to the neighborhood of its generator (intra-neighborhood references). The remaining 15 references are made to nodes other than the generator and its neighborhood (extra-neighborhood references). Where the neighborhood is less than the number of assigned intra-neighborhood references, the difference is carried over into the extra-neighborhood quota. 

For both intra and extra-neighborhood references, an agent generates a composite score for each potential target node and selects its citations by performing a weighted random sampling using the A-res algorithm presented in \cite{efraimidis2006weighted}. The process ensures that nodes with higher scores are more likely to be selected while maintaining stochasticity. Once an agent exhausts its quota of references, it does not make any more citations and can only receive citations from other agents. 

\subsection{Software}

Simulation code used in this study was implemented in Python 3.10, utilizing NetworKit 11.0 \citep{nwkt2014} for graph manipulation and multithreading with Numba 0.61.0 \citep{Lam2015} to enable parallelization. All data shown in this study were generated from experiments performed on the Illinois Campus Cluster with 16 cores of parallelism and varying memory configurations up to 128GB of RAM. Typical runtime was 45-60 minutes per simulation for a starting graph of $\sim$500,000 nodes and 30 cycles of simulation during which new agents were added in each cycle amounting to 3\% of the existing graph.  Simulations were tested at the 1\%, 3\% and 5\% levels and showed monotonic increases in output without any qualitative differences emerging at higher growth rates. The software, in its present form, does not permit a growth rate of 10\% under our standard conditions, even when memory is increased to 1TB.  The implementation is available as open source through Github \citep{abmcit2025}.

In the selection process, candidate nodes are split into two groups: nodes within the generator node’s neighborhood and the rest of the nodes in the existing network. A new node’s $\alpha$ value determines the ratio of citations that stay within the generator node’s neighborhood.  Preferential attachment and recency scores are computed through the following equations where $V$ and $E$ are the number of nodes and edges in the existing network, and $V_0$ is the set of vertices with no incoming edges.
$$Score_P(v_i)=\frac{\mbox{in-deg}(v_{i})^{\gamma{}} + C}{\sum_{v_{i} \in V}{\mbox{in-deg}(v_{i})^{\gamma{}} + C}} \mbox{where } \gamma{} \mbox{ and } C \mbox{ are set to 3 and 1 respectively}$$
$$Score_R(v_i)=\frac{\mbox{RecencyTable}[year_{current} - year_{i}] \cdot |{v_{j} \in V ; year_{j} = year_{i}|}}{\sum_{\mbox{each unique year of } year_{j} \in V }{\mbox{RecencyTable}[year_{current} - year_{j}]}}$$

Fitness values are assigned to both seed and agent nodes and are sampled from a power law distribution whose parameters were derived by obtaining a best-fit powerlaw exponent on the TCEN described in the Data subsection. The probability of a node receiving a fitness value of $x$ is $S \cdot C \cdot x^{E}$ where $S=6.37429$, $C=0.072$, $E=-1.634$.  A default fitness range of 1 to 1,000 was experimentally selected and is applied to all nodes with the exception of a single ultra-high fitness outlier whose chance of being created is 1 - cumulative probability from 1 to 1000. A user can suppress this high fitness single node and in such a case, the scale factor $S$ is adjusted accordingly such that the cumulative probability from 1 to 1000 equals 1. A user may also choose to specify additional nodes with specific fitness values beyond 1,000. $Score_F(v_i)$ is computed in the same way as $Score_P(v_i)$ except instead of in-degree we use the fitness values.

Finally, the weights for target nodes are given by a weighted linear combination of the scores and weights toward each score where the weights sum to 1.
$$Score_{Total}(v_i) = \emph{pw}*Score_P(v_i)+\emph{rw}*Score_R(v_i)+\emph{fw}*Score_F(v_i)$$ Thus, an agent's phenotype could comprise $w_P=0.3$, $w_R=0.4$,  $w_F=0.3$, and $\alpha=0.5$ with preferential weight (\emph{pw}), recency weight (\emph{rw}), and fitness weight (\emph{fw}).

\subsection{Terminology}
\begin{itemize}
\item \emph{in\_degree} interchangeable with count of citations received. 
\item \emph{out\_degree} interchangeably used with reference count
\item \emph{seed}: A node belonging to a real-world or random graph that is used as a starting point for a simulation. A seed node can be cited by an agent but cannot cite another node. A seed can have existing in\_degree.
\item \emph{agent}: A node in a real-world or random graph that is created during a simulation. An agent is initialized with zero in\_degree, cites other nodes and receives citations from other agents.
\item \emph{static} A homogeneous agent environment where all the agents in a simulation have the same citation phenotype, e.g.,  preferential weight (\emph{pw})=$0.3$, recency weight (\emph{rw})=$0.4$,  fitness weight (\emph{fw})=$0.3$, alpha ($\alpha$)=$0.5$
\item \emph{random} A heterogeneous agent environment where the agents in a simulation have randomly assigned phenotypes
\item \emph{hybrid} A partially homogeneous  agent environment where agents in a simulation have some fixed and some randomly assigned phenotype parameters, e.g.,  $\alpha=0.5$ and  pw, rw, and fw are randomly assigned. 
\item \emph{superstar}: An agent with fitness $>=$ 10,000 specified by the user. The default superstar setting is to have one superstar each of fitness 10,000, 100,000, and 1,000,000.
\end{itemize}

\subsection{Data} Citation data used in this study were derived from \cite{Wedell2022}, which focused on community structure in the exosome and extracellular vesicle  literature \citep{Raposo2023}, effectively initiated by two articles published in 1983 and experiencing spectacular growth after 2000. Briefly, the network described in \cite{Wedell2022} was curated to remove high-referencing articles and temporal inconsistencies such as an earlier article publishing a later article. The resultant graph named the Temporally Consistent Exosome Network (TCEN) was used in developing and testing the ABM models described below and consists of 13,983,371 nodes and 92,011,060 edges. A `seed' set used to initiate agent behavior consisted of all nodes in the TCEN published in 1982 or earlier as well as edges to other nodes from the time period. This seed set is referred to as the Stahl-Johnstone set and is named after the senior authors of two founder articles on exosomes published in 1983 \cite{Harding1983,Pan1983}. This seed set consists of 491,532 nodes and 899,050 edges with in\_degree of nodes (in-network citations) varying from 0-88435 and a median in\_degree of 1. One reason for the relatively low median in\_degree is the seed set expansion approach used to construct the network in \cite{Wedell2022} results in the outer layers having fewer citations or references.

For comparison, three Erd\H{o}s-R\'{e}nyi graphs were generated with the same number of nodes and edges as input to the `sample\_gnm' function in the igraph package for R  \citep{igraph}. 

\subsection{Out\_degree Distributions} Minimum and and maximum values were drawn from the \emph{sj} dataset, which ranged from 5 to 249 and were used to construct normal, power-law, and uniform distributions with integer values. For normally distributed out\_degree, the mean was set to 127 (max + min) / 2 and the standard deviation was set to 40. For power law, \emph{scipy.stats.powerlaw} function from the SciPy library \citep{2020SciPy-NMeth} was used with an exponent of 3 along with the  minimum and maximum values to generate 10,000 values. For a uniform distribution, the probability of any value was 1/(max- min + 1). 

\section{Summary ODD}
A complete, detailed model description, following the ODD (Overview, Design concepts, Details) protocol \citep{grimm2006, grimm2020} is provided at the end of the manuscript.

The purpose of our model is to understand how the growth of a citation network, scientific articles connected by citations, is influenced by citation patterns. The model includes the environment entity which represents the citation network and the agents which are the main actors creating citations. The most important process of the model invokes the ``cite'' submodel in which a new agent creates edges to existing agents in the network. In this submodel, every existing agent in the network receive a score based on their attributes, and a weighted random sampling is performed to select the nodes that receive citations. The exact details for the scoring is described in the software section of this manuscript as well as in the full ODD.

Initializing the model requires an \textit{edge\_list} defining the initial network structure and a \textit{node\_list} providing the \textit{year} attribute for each agent in the initial network structure. From then on, the model is driven by the attributes of the \textit{environment} such as the distributions used to sample the out-degrees or recency likelihoods along with the distributions used to sample agent weights and fitness values in the model. We describe the distributions used for this study under the Data section of this manuscript.

\section{Acknowledgments}

We thank Jay Mukherjee (Purdue University, West Lafayette, IN) for reviewing the code used in this project and providing helpful suggestions to improve it. We thank Jo\~{a}o Alfredo Cardoso Lamy (Insper Institute, Sao Paulo, Brazil) for generating scripts to analyze the output from simulations. The work done on this project was partly supported by an award from the Illinois-Insper Partnership. This research was supported in part by the Illinois Computes project, which is supported by the University of Illinois Urbana-Champaign and the University of Illinois System. We apologize to those whose work we did not cite and discuss.


%



% ...




 
\bibliography{abm_arxiv.bib} 

\begin{thebibliography}{55}
\providecommand{\natexlab}[1]{#1}
\providecommand{\url}[1]{\texttt{#1}}
\expandafter\ifx\csname urlstyle\endcsname\relax
  \providecommand{\doi}[1]{doi: #1}\else
  \providecommand{\doi}{doi: \begingroup \urlstyle{rm}\Url}\fi

\bibitem[Price(1965)]{Price1965}
Derek J. de~Solla Price.
\newblock Networks of scientific papers: The pattern of bibliographic
  references indicates the nature of the scientific research front.
\newblock \emph{Science}, 149\penalty0 (3683):\penalty0 510–515, July 1965.
\newblock ISSN 1095-9203.
\newblock \doi{10.1126/science.149.3683.510}.
\newblock URL \url{http://dx.doi.org/10.1126/science.149.3683.510}.

\bibitem[Chubin and Moitra(1975)]{Chubin1975}
Daryl~E. Chubin and Soumyo~D. Moitra.
\newblock Content analysis of references: Adjunct or alternative to citation
  counting?
\newblock \emph{Social Studies of Science}, 5\penalty0 (4):\penalty0 423–441,
  November 1975.
\newblock ISSN 1460-3659.
\newblock \doi{10.1177/030631277500500403}.
\newblock URL \url{http://dx.doi.org/10.1177/030631277500500403}.

\bibitem[Moravcsik and Murugesan(1975)]{Moravcsik1975}
Michael~J. Moravcsik and Poovanalingam Murugesan.
\newblock Some results on the function and quality of citations.
\newblock \emph{Social Studies of Science}, 5\penalty0 (1):\penalty0 86–92,
  February 1975.
\newblock ISSN 1460-3659.
\newblock \doi{10.1177/030631277500500106}.
\newblock URL \url{http://dx.doi.org/10.1177/030631277500500106}.

\bibitem[Price(1976)]{Price1976}
Derek De~Solla Price.
\newblock A general theory of bibliometric and other cumulative advantage
  processes.
\newblock \emph{Journal of the American Society for Information Science},
  27\penalty0 (5):\penalty0 292–306, September 1976.
\newblock ISSN 1097-4571.
\newblock \doi{10.1002/asi.4630270505}.
\newblock URL \url{http://dx.doi.org/10.1002/asi.4630270505}.

\bibitem[Gilbert(1977)]{Gilbert1977}
G.~Nigel Gilbert.
\newblock Referencing as persuasion.
\newblock \emph{Social Studies of Science}, 7\penalty0 (1):\penalty0 113–122,
  February 1977.
\newblock ISSN 1460-3659.
\newblock \doi{10.1177/030631277700700112}.
\newblock URL \url{http://dx.doi.org/10.1177/030631277700700112}.

\bibitem[Small(1978)]{Small1978}
Henry~G. Small.
\newblock Cited documents as concept symbols.
\newblock \emph{Social Studies of Science}, 8\penalty0 (3):\penalty0 327–340,
  August 1978.
\newblock ISSN 1460-3659.
\newblock \doi{10.1177/030631277800800305}.
\newblock URL \url{http://dx.doi.org/10.1177/030631277800800305}.

\bibitem[Cozzens(1981)]{Cozzens1981}
Susan~E. Cozzens.
\newblock Taking the measure of science: A review of citation theories.
\newblock \emph{Newsletter of the International Society for the Sociology of
  Knowledge}, 7\penalty0 (1–2):\penalty0 16--21, May 1981.

\bibitem[Zuckerman(1987)]{Zuckerman1987}
Harriet Zuckerman.
\newblock Citation analysis and the complex problem of intellectual influence.
\newblock \emph{Scientometrics}, 12\penalty0 (5–6):\penalty0 329–338,
  November 1987.
\newblock ISSN 1588-2861.
\newblock \doi{10.1007/bf02016675}.
\newblock URL \url{http://dx.doi.org/10.1007/BF02016675}.

\bibitem[Martin and Irvine(1983)]{Martin1983}
Ben~R. Martin and John Irvine.
\newblock Assessing basic research.
\newblock \emph{Research Policy}, 12\penalty0 (2):\penalty0 61–90, April
  1983.
\newblock ISSN 0048-7333.
\newblock \doi{10.1016/0048-7333(83)90005-7}.
\newblock URL \url{http://dx.doi.org/10.1016/0048-7333(83)90005-7}.

\bibitem[MacRoberts and MacRoberts(1996)]{MacRoberts1996}
M.~H. MacRoberts and Barbara~R. MacRoberts.
\newblock Problems of citation analysis.
\newblock \emph{Scientometrics}, 36\penalty0 (3):\penalty0 435–444, July
  1996.
\newblock ISSN 1588-2861.
\newblock \doi{10.1007/bf02129604}.
\newblock URL \url{http://dx.doi.org/10.1007/BF02129604}.

\bibitem[Stigler(1987)]{Stigler8bba9ab5c485}
Stephen~M. Stigler.
\newblock {Precise Measurement in the Face of Error: A Comment on MacRoberts
  and MacRoberts}.
\newblock \emph{Social Studies of Science}, 17\penalty0 (2):\penalty0 332--334,
  1987.
\newblock ISSN 03063127.
\newblock URL \url{http://www.jstor.org/stable/284955}.

\bibitem[Leydesdorff(1998)]{Leydesdorff1998}
L.~Leydesdorff.
\newblock Theories of citation?
\newblock \emph{Scientometrics}, 43\penalty0 (1):\penalty0 5–25, September
  1998.
\newblock ISSN 1588-2861.
\newblock \doi{10.1007/bf02458391}.
\newblock URL \url{http://dx.doi.org/10.1007/BF02458391}.

\bibitem[Wouters(1999)]{Wouters1999TheCC}
Paul Wouters.
\newblock \emph{The citation culture}.
\newblock PhD thesis, University of Amsterdam, 1999.

\bibitem[Aksnes et~al.(2019)Aksnes, Langfeldt, and Wouters]{Aksnes2019}
Dag~W. Aksnes, Liv Langfeldt, and Paul Wouters.
\newblock Citations, citation indicators, and research quality: An overview of
  basic concepts and theories.
\newblock \emph{Sage Open}, 9\penalty0 (1), 2019.
\newblock ISSN 2158-2440.
\newblock \doi{10.1177/2158244019829575}.
\newblock URL \url{http://dx.doi.org/10.1177/2158244019829575}.

\bibitem[Olszewski(2020)]{Olszewski2020}
Wojciech Olszewski.
\newblock A theory of citations.
\newblock \emph{Research in Economics}, 74\penalty0 (3):\penalty0 193–212,
  September 2020.
\newblock ISSN 1090-9443.
\newblock \doi{10.1016/j.rie.2020.06.001}.
\newblock URL \url{http://dx.doi.org/10.1016/j.rie.2020.06.001}.

\bibitem[Cronin(1981)]{Cronin1981TheNF}
Blaise Cronin.
\newblock The need for a theory of citing.
\newblock \emph{J. Documentation}, 37:\penalty0 16--24, 1981.
\newblock URL \url{https://api.semanticscholar.org/CorpusID:42485292}.

\bibitem[Cozzens(1989)]{Cozzens1989}
Susan~E. Cozzens.
\newblock What do citations count? the rhetoric-first model.
\newblock \emph{Scientometrics}, 15\penalty0 (5–6):\penalty0 437–447, May
  1989.
\newblock ISSN 1588-2861.
\newblock \doi{10.1007/bf02017064}.
\newblock URL \url{http://dx.doi.org/10.1007/bf02017064}.

\bibitem[Merton(1968)]{Merton1968}
Robert~K. Merton.
\newblock The matthew effect in science: The reward and communication systems
  of science are considered.
\newblock \emph{Science}, 159\penalty0 (3810):\penalty0 56–63, January 1968.
\newblock ISSN 1095-9203.
\newblock \doi{10.1126/science.159.3810.56}.
\newblock URL \url{http://dx.doi.org/10.1126/science.159.3810.56}.

\bibitem[Latour and Woolgar(1986)]{Latour1986-sy}
Bruno Latour and Steve Woolgar.
\newblock \emph{Laboratory life}.
\newblock Princeton University Press, Princeton, NJ, September 1986.

\bibitem[Leopold(1973)]{Leopold1973}
A.~C. Leopold.
\newblock Games scientists play.
\newblock \emph{BioScience}, 23\penalty0 (10):\penalty0 590–594, October
  1973.
\newblock ISSN 1525-3244.
\newblock \doi{10.2307/1296499}.
\newblock URL \url{http://dx.doi.org/10.2307/1296499}.

\bibitem[Hacking(1994)]{Hacking1994}
Ian Hacking.
\newblock The advancement of science: Science without legend, objectivity
  without illusion by philip kitcher.
\newblock \emph{Journal of Philosophy}, 91\penalty0 (4):\penalty0 212--215,
  1994.
\newblock \doi{10.5840/jphil199491435}.

\bibitem[Kitcher(1993)]{Kitcher1993-KITTAO-2}
Philip Kitcher.
\newblock \emph{The Advancement of Science: Science Without Legend, Objectivity
  Without Illusions}.
\newblock Oxford University Press, New York, 1993.

\bibitem[Watts and Gilbert(2011)]{Watts2011}
Christopher Watts and Nigel Gilbert.
\newblock Does cumulative advantage affect collective learning in science? an
  agent-based simulation.
\newblock \emph{Scientometrics}, 89\penalty0 (1):\penalty0 437–463, June
  2011.
\newblock ISSN 1588-2861.
\newblock \doi{10.1007/s11192-011-0432-8}.
\newblock URL \url{http://dx.doi.org/10.1007/s11192-011-0432-8}.

\bibitem[Kleinberg et~al.(1999)Kleinberg, Kumar, Raghavan, Rajagopalan, and
  Tomkins]{Kleinberg1999}
Jon~M. Kleinberg, Ravi Kumar, Prabhakar Raghavan, Sridhar Rajagopalan, and
  Andrew~S. Tomkins.
\newblock \emph{The Web as a Graph: Measurements, Models, and Methods}, page
  1–17.
\newblock Springer Berlin Heidelberg, 1999.
\newblock ISBN 9783540486862.
\newblock \doi{10.1007/3-540-48686-0\_1}.
\newblock URL \url{http://dx.doi.org/10.1007/3-540-48686-0_1}.

\bibitem[Krapivsky and Redner(2005)]{Krapivsky2005}
P.~L. Krapivsky and S.~Redner.
\newblock Network growth by copying.
\newblock \emph{Physical Review E}, 71\penalty0 (3), March 2005.
\newblock ISSN 1550-2376.
\newblock \doi{10.1103/physreve.71.036118}.
\newblock URL \url{http://dx.doi.org/10.1103/PhysRevE.71.036118}.

\bibitem[Touwen et~al.(2024)Touwen, Bucur, van~der Hofstad, Garavaglia, and
  Litvak]{Touwen2024}
Lourens Touwen, Doina Bucur, Remco van~der Hofstad, Alessandro Garavaglia, and
  Nelly Litvak.
\newblock Learning the mechanisms of network growth.
\newblock \emph{Scientific Reports}, 14\penalty0 (1), May 2024.
\newblock ISSN 2045-2322.
\newblock \doi{10.1038/s41598-024-61940-4}.
\newblock URL \url{http://dx.doi.org/10.1038/s41598-024-61940-4}.

\bibitem[Zhou et~al.(2023)Zhou, Holme, Gong, Zhan, Huang, Lu, and
  Meng]{Zhou2023}
Bin Zhou, Petter Holme, Zaiwu Gong, Choujun Zhan, Yao Huang, Xin Lu, and
  Xiangyi Meng.
\newblock The nature and nurture of network evolution.
\newblock \emph{Nature Communications}, 14\penalty0 (1), November 2023.
\newblock ISSN 2041-1723.
\newblock \doi{10.1038/s41467-023-42856-5}.
\newblock URL \url{http://dx.doi.org/10.1038/s41467-023-42856-5}.

\bibitem[Simon(1955)]{simon1955class}
Herbert~A Simon.
\newblock On a class of skew distribution functions.
\newblock \emph{Biometrika}, 42\penalty0 (3/4):\penalty0 425--440, 1955.

\bibitem[Barab{\'a}si and Albert(1999)]{barabasi1999emergence}
Albert-L{\'a}szl{\'o} Barab{\'a}si and R{\'e}ka Albert.
\newblock Emergence of scaling in random networks.
\newblock \emph{science}, 286\penalty0 (5439):\penalty0 509--512, 1999.

\bibitem[Silva et~al.(2020)Silva, Tandon, Amancio, Flammini, Menczer,
  Milojević, and Fortunato]{Silva2020}
Filipi~Nascimento Silva, Aditya Tandon, Diego~Raphael Amancio, Alessandro
  Flammini, Filippo Menczer, Staša Milojević, and Santo Fortunato.
\newblock Recency predicts bursts in the evolution of author citations.
\newblock \emph{Quantitative Science Studies}, 1\penalty0 (3):\penalty0
  1298–1308, August 2020.
\newblock ISSN 2641-3337.
\newblock \doi{10.1162/qss\_a\_00070}.
\newblock URL \url{http://dx.doi.org/10.1162/qss\_a\_00070}.

\bibitem[Gilbert(1997)]{Gilbert1997}
N.~Gilbert.
\newblock A simulation of the structure of academic science.
\newblock \emph{Sociological Research Online}, 2\penalty0 (2):\penalty0
  91–105, June 1997.
\newblock ISSN 1360-7804.
\newblock \doi{10.5153/sro.85}.
\newblock URL \url{http://dx.doi.org/10.5153/sro.85}.

\bibitem[Wedell et~al.(2022)Wedell, Park, Korobskiy, Warnow, and
  Chacko]{Wedell2022}
Eleanor Wedell, Minhyuk Park, Dmitriy Korobskiy, Tandy Warnow, and George
  Chacko.
\newblock Center–periphery structure in research communities.
\newblock \emph{Quantitative Science Studies}, 3\penalty0 (1):\penalty0
  289–314, 2022.
\newblock ISSN 2641-3337.
\newblock \doi{{10.1162/qss\_a\_00184}}.
\newblock URL \url{http://dx.doi.org/10.1162/qss\_a\_00184}.

\bibitem[B\"{o}rner et~al.(2004)B\"{o}rner, Maru, and Goldstone]{Borner2004}
Katy B\"{o}rner, Jeegar~T. Maru, and Robert~L. Goldstone.
\newblock The simultaneous evolution of author and paper networks.
\newblock \emph{Proceedings of the National Academy of Sciences}, 101\penalty0
  (suppl\_1):\penalty0 5266–5273, April 2004.
\newblock ISSN 1091-6490.
\newblock \doi{10.1073/pnas.0307625100}.
\newblock URL \url{http://dx.doi.org/10.1073/pnas.0307625100}.

\bibitem[Shah et~al.(2019)Shah, Kumar, and Sundaram]{Shah2019}
Harshay Shah, Suhansanu Kumar, and Hari Sundaram.
\newblock Growing attributed networks through local processes.
\newblock In \emph{The World Wide Web Conference}, WWW ’19, page 3208–3214.
  ACM, May 2019.
\newblock \doi{10.1145/3308558.3313640}.
\newblock URL \url{http://dx.doi.org/10.1145/3308558.3313640}.

\bibitem[Weisberg and Muldoon(2009)]{Weisberg2009}
Michael Weisberg and Ryan Muldoon.
\newblock Epistemic landscapes and the division of cognitive labor.
\newblock \emph{Philosophy of Science}, 76\penalty0 (2):\penalty0 225–252,
  April 2009.
\newblock ISSN 1539-767X.
\newblock \doi{10.1086/644786}.
\newblock URL \url{http://dx.doi.org/10.1086/644786}.

\bibitem[Harnagel(2019)]{Harnagel2019}
Audrey Harnagel.
\newblock A mid-level approach to modeling scientific communities.
\newblock \emph{Studies in History and Philosophy of Science Part A},
  76:\penalty0 49–59, August 2019.
\newblock ISSN 0039-3681.
\newblock \doi{10.1016/j.shpsa.2018.12.010}.
\newblock URL \url{http://dx.doi.org/10.1016/j.shpsa.2018.12.010}.

\bibitem[Alexandrova(2017)]{Alexandrova2017}
Anna Alexandrova.
\newblock \emph{A Philosophy for the Science of Well-Being}.
\newblock Oxford University Press, New York, 2017.

\bibitem[Raposo and Stahl(2023)]{Raposo2023}
Graca Raposo and Philip~D. Stahl.
\newblock Extracellular vesicles - on the cusp of a new language in the
  biological sciences.
\newblock \emph{Extracellular Vesicles and Circulating Nucleic Acids},
  4\penalty0 (2):\penalty0 240–54, 2023.
\newblock ISSN 2767-6641.
\newblock \doi{10.20517/evcna.2023.18}.

\bibitem[Lowry et~al.(1951)Lowry, Rosebrough, Farr, and Randall]{Lowry1951}
OliverH. Lowry, NiraJ. Rosebrough, A.~Lewis Farr, and RoseJ. Randall.
\newblock Protein measurement with the folin phenol reagent.
\newblock \emph{Journal of Biological Chemistry}, 193\penalty0 (1):\penalty0
  265–275, November 1951.
\newblock ISSN 0021-9258.
\newblock \doi{10.1016/s0021-9258(19)52451-6}.
\newblock URL \url{http://dx.doi.org/10.1016/s0021-9258(19)52451-6}.

\bibitem[Van~Noorden et~al.(2014)Van~Noorden, Maher, and Nuzzo]{VanNoorden2014}
Richard Van~Noorden, Brendan Maher, and Regina Nuzzo.
\newblock The top 100 papers.
\newblock \emph{Nature}, 514\penalty0 (7524):\penalty0 550–553, October 2014.
\newblock ISSN 1476-4687.
\newblock \doi{10.1038/514550a}.
\newblock URL \url{http://dx.doi.org/10.1038/514550a}.

\bibitem[De~Solla~Price and Beaver(1966)]{DeSollaPrice1966}
Derek~J. De~Solla~Price and Donald Beaver.
\newblock Collaboration in an invisible college.
\newblock \emph{American Psychologist}, 21\penalty0 (11):\penalty0 1011–1018,
  1966.
\newblock ISSN 0003-066X.
\newblock \doi{10.1037/h0024051}.
\newblock URL \url{http://dx.doi.org/10.1037/h0024051}.

\bibitem[Small and Sweeney(1985)]{Small1985sw}
H.~Small and E.~Sweeney.
\newblock Clustering the science citation index {\textregistered} using
  co-citations: I. a comparison of methods.
\newblock \emph{Scientometrics}, 7\penalty0 (3–6):\penalty0 391–409, March
  1985.
\newblock ISSN 1588-2861.
\newblock \doi{10.1007/bf02017157}.
\newblock URL \url{http://dx.doi.org/10.1007/BF02017157}.

\bibitem[Leydesdorff and Opthof(2010)]{Leydesdorff2010}
Loet Leydesdorff and Tobias Opthof.
\newblock Scopus’s source normalized impact per paper (snip) versus a journal
  impact factor based on fractional counting of citations.
\newblock \emph{Journal of the American Society for Information Science and
  Technology}, 61\penalty0 (11):\penalty0 2365–2369, May 2010.
\newblock ISSN 1532-2890.
\newblock \doi{10.1002/asi.21371}.
\newblock URL \url{http://dx.doi.org/10.1002/asi.21371}.

\bibitem[Castillo-Castillo et~al.(2025)Castillo-Castillo, Stevens-Navarro,
  Pineda-Rico, Garcia-Barrientos, Castillo-Soria, and
  Acosta-Elias]{Castillo-Castillo2025-ov}
Pedro Castillo-Castillo, Enrique Stevens-Navarro, Ulises Pineda-Rico, Abel
  Garcia-Barrientos, Francisco~R Castillo-Soria, and Jesus Acosta-Elias.
\newblock A growth model for citations networks.
\newblock \emph{Appl. Netw. Sci.}, 10\penalty0 (1), jan 2025.
\newblock \doi{10.1007/s41109‑025‑00691‑1}.

\bibitem[Biagioli and Lippman(2020)]{biagioli2020}
Mario Biagioli and Alexandra Lippman.
\newblock \emph{Gaming the Metrics: Misconduct and Manipulation in Academic
  Research}.
\newblock The MIT Press, 2020.
\newblock ISBN 9780262356565.
\newblock \doi{10.7551/mitpress/11087.001.0001}.
\newblock URL \url{http://dx.doi.org/10.7551/mitpress/11087.001.0001}.

\bibitem[Efraimidis and Spirakis(2006)]{efraimidis2006weighted}
Pavlos~S Efraimidis and Paul~G Spirakis.
\newblock Weighted random sampling with a reservoir.
\newblock \emph{Information processing letters}, 97\penalty0 (5):\penalty0
  181--185, 2006.

\bibitem[Staudt et~al.(2014)Staudt, Sazonovs, and Meyerhenke]{nwkt2014}
Christian~L. Staudt, Aleksejs Sazonovs, and Henning Meyerhenke.
\newblock Networkit: A tool suite for large-scale complex network analysis,
  2014.
\newblock URL \url{https://arxiv.org/abs/1403.3005}.

\bibitem[Lam et~al.(2015)Lam, Pitrou, and Seibert]{Lam2015}
Siu~Kwan Lam, Antoine Pitrou, and Stanley Seibert.
\newblock Numba: a llvm-based python jit compiler.
\newblock In \emph{Proceedings of the Second Workshop on the LLVM Compiler
  Infrastructure in HPC}, SC15, page 1–6. ACM, November 2015.
\newblock \doi{10.1145/2833157.2833162}.
\newblock URL \url{http://dx.doi.org/10.1145/2833157.2833162}.

\bibitem[Ramavarapu et~al.(2025)Ramavarapu, Park, {Robles Granda}, Warnow, and
  Chacko]{abmcit2025}
Vikram Ramavarapu, Minhyuk Park, Pablo {Robles Granda}, Tandy Warnow, and
  George Chacko.
\newblock abm\_citations (version 3), 2025.
\newblock URL
  \url{https://github.com/illinois-or-research-analytics/abm_citations}.

\bibitem[Harding et~al.(1983)Harding, Heuser, and Stahl]{Harding1983}
C~Harding, J~Heuser, and P~Stahl.
\newblock Receptor-mediated endocytosis of transferrin and recycling of the
  transferrin receptor in rat reticulocytes.
\newblock \emph{The Journal of cell biology}, 97\penalty0 (2):\penalty0
  329–339, August 1983.
\newblock ISSN 1540-8140.
\newblock \doi{10.1083/jcb.97.2.329}.
\newblock URL \url{http://dx.doi.org/10.1083/jcb.97.2.329}.

\bibitem[Pan and Johnstone(1983)]{Pan1983}
Bin-Tao Pan and Rose~M. Johnstone.
\newblock Fate of the transferrin receptor during maturation of sheep
  reticulocytes in vitro: Selective externalization of the receptor.
\newblock \emph{Cell}, 33\penalty0 (3):\penalty0 967–978, July 1983.
\newblock ISSN 0092-8674.
\newblock \doi{10.1016/0092-8674(83)90040-5}.
\newblock URL \url{http://dx.doi.org/10.1016/0092-8674(83)90040-5}.

\bibitem[Csardi and Nepusz(2006)]{igraph}
Gabor Csardi and Tamas Nepusz.
\newblock The igraph software package for complex network research.
\newblock \emph{InterJournal}, Complex Systems:\penalty0 1695, 2006.
\newblock URL \url{https://igraph.org}.

\bibitem[Virtanen et~al.(2020)Virtanen, Gommers, Oliphant, Haberland, Reddy,
  Cournapeau, Burovski, Peterson, Weckesser, Bright, {van der Walt}, Brett,
  Wilson, Millman, Mayorov, Nelson, Jones, Kern, Larson, Carey, Polat, Feng,
  Moore, {VanderPlas}, Laxalde, Perktold, Cimrman, Henriksen, Quintero, Harris,
  Archibald, Ribeiro, Pedregosa, {van Mulbregt}, and {SciPy 1.0
  Contributors}]{2020SciPy-NMeth}
Pauli Virtanen, Ralf Gommers, Travis~E. Oliphant, Matt Haberland, Tyler Reddy,
  David Cournapeau, Evgeni Burovski, Pearu Peterson, Warren Weckesser, Jonathan
  Bright, St{\'e}fan~J. {van der Walt}, Matthew Brett, Joshua Wilson, K.~Jarrod
  Millman, Nikolay Mayorov, Andrew R.~J. Nelson, Eric Jones, Robert Kern, Eric
  Larson, C~J Carey, {\.I}lhan Polat, Yu~Feng, Eric~W. Moore, Jake
  {VanderPlas}, Denis Laxalde, Josef Perktold, Robert Cimrman, Ian Henriksen,
  E.~A. Quintero, Charles~R. Harris, Anne~M. Archibald, Ant{\^o}nio~H. Ribeiro,
  Fabian Pedregosa, Paul {van Mulbregt}, and {SciPy 1.0 Contributors}.
\newblock {{SciPy} 1.0: Fundamental Algorithms for Scientific Computing in
  Python}.
\newblock \emph{Nature Methods}, 17:\penalty0 261--272, 2020.
\newblock \doi{10.1038/s41592-019-0686-2}.

\bibitem[Grimm et~al.(2006)Grimm, Berger, Bastiansen, Eliassen, Ginot, Giske,
  Goss-Custard, Grand, Heinz, Huse, et~al.]{grimm2006}
Volker Grimm, Uta Berger, Finn Bastiansen, Sigrunn Eliassen, Vincent Ginot,
  Jarl Giske, John Goss-Custard, Tamara Grand, Simone~K Heinz, Geir Huse,
  et~al.
\newblock A standard protocol for describing individual-based and agent-based
  models.
\newblock \emph{Ecological modelling}, 198\penalty0 (1-2):\penalty0 115--126,
  2006.

\bibitem[Grimm et~al.(2020)Grimm, Railsback, Vincenot, Berger, Gallagher,
  DeAngelis, Edmonds, Ge, Giske, Groeneveld, et~al.]{grimm2020}
Volker Grimm, Steven~F Railsback, Christian~E Vincenot, Uta Berger, Cara
  Gallagher, Donald~L DeAngelis, Bruce Edmonds, Jiaqi Ge, Jarl Giske, Juergen
  Groeneveld, et~al.
\newblock The odd protocol for describing agent-based and other simulation
  models: A second update to improve clarity, replication, and structural
  realism.
\newblock \emph{Journal of Artificial Societies and Social Simulation},
  23\penalty0 (2), 2020.

\end{thebibliography}
\section{Model}
The model description follows the ODD (Overview, Design concepts, Details) protocol for describing individual- and agent-based models \citep{grimm2006}, as updated by \citet{grimm2020}.

\subsection{Purpose and patterns}
The purpose of this model is to understand how the growth of a citation network, scientific articles connected by citations, is influenced by citation patterns. A distal question is whether citation strategy can reward a researcher and an overarching problem is to quantitatively elucidate extant theories of citation that have been presented qualitatively.To this purpose, we simulate the growth of a citation network under different conditions that address preferential attachment, recency, citation fitness, and epistemic proximity.

\begin{itemize}
    \item \textbf{Pattern 1: Degree distribution.} As bias towards preferential attachment increases in an agent, a corresponding increase in\_degree (citations) occurs.
    \item \textbf{Pattern 2: Degree distribution.} As node fitness increases, a corresponding increase in\_degree (citations) occurs.
\end{itemize}

\subsection{Entities, state variables, and scales}
This model consist of two entities: the agents (publications that cite other publications) and the environment (the citation network created by the agents). The state variables for each agent are described in table \ref{tab:agent-state-variables}, and the state variables for the environment are described in table \ref{tab:environment-state-variables}.

Agents
\textbf{Rationale:}  Mirroring the growth of the scientific literature, agents are created in time-stamped batches to represent new publications in a growing citation network. Each agent makes citations to other agents (references or out\_degree) and, in turn, receives citations (in\_degree) from other agents create later in time. Each agent has a unique identifier and agents vary according to the year in which they were published (recency), their citation fitness, and their bias toward epistemic proximity (network neighbors), recency, and preferential attachment (affinity for higher in\_degree) of their citation targets.

Environment
\textbf{Rationale:} The environment is the citation network; a representation of a complex system where agents live in the topological space represented of a  directed graph with unweighted edges. The environment defines parameters for growth rate of the network.

\begin{table}[!t]
\centering
\begin{tabular}{lll}
\hline
Variable name & Variable characteristics & Meaning \\
\hline
node\_id & static; +ve integer value & unique identifier for agent \\
pub\_year & static; +ve integer value & represents the year in which agent was initialized (publication year) \\
pa\_weight & static; numeric value [0, 1] & influence of preferential attachment \\
recency\_weight & static; numeric value [0, 1] & influence of recency  \\
fitness\_weight & static; numeric value [0, 1]& influence of fitness \\
alpha & static; numeric value [0, 1]& influence of epistemic proximity \\
num\_citations & static; +ve integer value & number of citations that this agent make \\
fitness & static; +ve integer value & the quality of an agent to attract citations \\
\hline
\end{tabular}
\label{tab:agent-state-variables}
\caption{Agent state variables}
\end{table}

\begin{table}[!t]
\centering
\begin{tabular}{lll}
\hline
Variable name & Variable characteristics & Meaning \\
\hline
edge\_list & dynamic & a directed graph of agents as nodes and citations as edges \\
same\_year\_percentage & static; numeric value [0, 1] & the fraction of agents that cite another agent from the same year \\
recency\_distribution & static; two-column integer list & probability of a citation based on the difference in agent years  \\
reference\_count\_distribution & static; +ve integer list & the distribution used to allot citations to agents \\
gamma & static; numeric value & the exponent term in preferential attachment \\
c & static; integer value & the constant term in preferential attachment \\
growth\_rate & static; numeric value & factor for growth \\
\hline
\end{tabular}
\label{tab:environment-state-variables}
\caption{Environment state variables}
\end{table}

The scale of our model is defined by three factors: initial network size, growth rate, and the number of years. The network growth follows an exponential growth formula where in each year, a percentage, defined by the growth rate, of the current network size determines the number of new agents that are initialized. The number of years then simply defines how many growth cycles the simulation undergoes. One year represents one cycle of agent initialization.

\subsection{Process overview and scheduling}
\textit{Processes}: Our model is designed to simulate the growth of a citation network starting with a base network and introducing new agents that make citations independently. There are four processes: two for agents (generator node selection and citation creation) and two for the environment (agent initilaization and environment update).
\begin{itemize}
    \item Agent initialization: Agent initilaization involves assigning where all agents needed for the current year are generated with the correct state variables (\textit{node\_id, pub\_year, pa\_weight, recency\_weight, fitness\_weight, num\_citations, fitness}).
    \item Environment update: When new nodes and edges are created, they are not added back into the environment untl all agents of a given year are created and edges made. Once all agents have finished making their citatitions, they are all added to the environment at once. Here, the \textit{edge\_list} attribute of the environment is updated.
    \item Generator node selection: An existing agent is randomly selected to be a generator node of another agent upon initialization. There is an edge created from an agent to its generator node.
    \item Citation creation: For a given set of candidate nodes, the ``cite'' submodel is used to rank and perform a weighted sampling of the nodes. Edges are created to these chosen nodes.
\end{itemize}

\textit{Schedule}:
\begin{enumerate}
    \item Initially, the environment is initilized as described in the Initialization section.
    \item For each year of simulation, the following steps are repeated
    \begin{enumerate}
        \item ``Agent initilaization'' process
        \item Edges are created for \textit{same\_year\_percentage} proportion of new agents to cite other new agents from the same year.
        \item For each node created:
        \begin{enumerate}
            \item ``Generator node selection'' process
            \item ``Citation creation'' process for 1-hop neighborhood of the generator node
            \item ``Citation creation'' process for outside the 1-hop neighborhood of the generator node
        \end{enumerate}
        \item ``Environment update'' process
    \end{enumerate}
\end{enumerate}

\subsection{Design concepts}
\subsubsection{Basic principles}
The model parsimoniously represents the growth of a citation graph based on known citation patterns and theories of network growth. In it, autonomous agents representing publications make citations to other agents to simulate the process of referencing other articles. New agents are added in batches corresponding to a year of new articles. The process is initiated using either a citation graph or a random network representing an existing epistemic landscape in a field of research.

\subsubsection{Emergence}
As each agent makes citations, the network grows and it is possible to evaluate the basis for accrued citations. The mechanism by which the agents make citations is driven by the ``Citation creation'' process which models a researcher evaluating their environment and making decisions based on the influences of preferential attachment, recency, and fitness.

\subsubsection{Adaptation}
Agents upon initialization are modeled via indirect objective-seeking as an agent makes citations based on the in-degrees, publication year, and fitness attributes of other agents combined with its own weights for each of those components. Agents are not adaptive after they are created. Once an agent's quota of citations is made, it only receive citations from subsequently created agents.

\subsubsection{Objectives}
The measure of success of each agent is its eventual in-degree at the end of the model simulation. In order to achieve this goal with limited \textit{out\_degree}, each agent uses the ``cite'' submodel to analyze the best way of allocating its limited resources.

\subsubsection{Learning}
Learning is not implemented and is deferred to a future version of this model.

\subsubsection{Prediction}
The adaptive behavior of agents at initilization is based on the implicit prediction that the chances of receiving citations in future cycles of the model will be impacted by the edges that each agent creates.

\subsubsection{Sensing}
Each agent is fully aware of all of the state variables for every other agent as well as the environment.

\subsubsection{Interaction}
Each agent interacts with another agent via citations. An agent citing another agent affects the \textit{edge\_list} state variable of the environment.

\subsubsection{Stochasticity}
The processes generator node selection and citation creation for the agents and agent initialization for the environment are stochastic as they inherently use randomness. The generator nodes for each agent are chosen randomly, the citations for each agent are chosen through a weighted random process, and each agent gets initialized with random weights.

\subsubsection{Collectives}
This model includes no collectives.

\subsubsection{Observation}
The main output of our model is the resulting citation network in the form of an edgelist, which is the result of agents citing other agents for the duration of the simulation. The year in which the citations are created are also recorded.

\subsection{Initialization}
The model is initiazilized with initial agents loaded from the \textit{edge\_list} and \textit{node\_list}, provided as input data. The initialization also includes a \textit{fitness} value. For the initial agents, unlike those agents created during the simulation, weights are not required to be initialized as the edges for these agents are created from the input \textit{edge\_list} rather than created through the weights.

The \textit{recency\_distribution} attribute of the environment is initilialized from an input table where the likelihood of citing an agent from $x$ years ago is provided. Therefore, the model can only simulate upto those years whose likelihoods are defined in the input table and thus present in the \textit{recency\_distribution} attribute.

The \textit{reference\_count\_distribution} attribute of the environment is initilialized from an input table where the likelihood of creating an agent with a certain out-degree is provided.

\textit{same\_year\_percentage}, \textit{gamma}, \textit{c}, \textit{growth\_rate}, and the number of years for which to run the model are to be provided by the user when starting the model.

The model is intended to be generic where users can specify different ways of initializing the input such as different \textit{edge\_list} or \textit{node\_list}. The model only requires that the agents initially are given the \textit{year} attribute and edges among them. Given the same input, the model will deterministically create the same topology for the initial network. Note that the model will not create the same final network even with the same initial network.

\subsection{Input data}
The input data required to run the model is an \textit{edge\_list} providing the necessary edge information for initial agents, \textit{node\_list} defining the \textit{year} attributes for each node in the initial seed network, and tables representing the distributions required to initialize the \textit{recency\_distribution} and \textit{reference\_count\_distribution}. These are used in the initialization step.

\subsection{Submodels}

\subsubsection{cite} The cite submodel works as follows.
Given a set of candidates nodes $V$ to which to make citations, the cite submodel does a weighted random selection of nodes where each node in $V$ has a score that serves as the weight in weighted random sampling. We show how to compute the score each node below.

$$Score_{Total}(v_i) = \emph{pw}*Score_P(v_i)+\emph{rw}*Score_R(v_i)+\emph{fw}*Score_F(v_i)$$ where \emph{pw}, \emph{rw}, and \emph{fw} are preferential, recency, and fitness weights, respectively.

The weights are proportional to different node properties where
$$Score_P(v_i) \sim \mbox{in-deg}(v_{i})^\gamma{} + C$$
$$Score_R(v_i) \sim \mbox{RecencyTable}[year_{current} - year_{i}] \cdot |{v_{j} \in V ; year_{j} = year_{i}|}$$
$$Score_F(v_i) \sim \mbox{fitness}(v_i)^\gamma + C$$

The $\gamma{}$, $C$, and $\mbox{RecencyTable}$ variables come from the attributes of the environment which are initilaized during the initialization step. Specifically, the $\mbox{RecencyTable}$ variable is a short-hand for accessing the \textit{recency\_distribution} attribute where the query is for the likelihood of citing an agent from $x$ years ago.

\end{document}